\documentclass[fleqn, usenatbib]{mnras}

\usepackage{newtxtext,newtxmath}
\usepackage[T1]{fontenc}
\usepackage{ae,aecompl}

\usepackage{graphicx}	        
\usepackage{amsmath}	        
\usepackage{url}
\usepackage{microtype}
\usepackage{xspace}
\usepackage{float}
\usepackage[usenames]{color}

\newcommand{\msun}{\ensuremath{\mathrm{M}_{\odot}}}   
\newcommand{\rsun}{\ensuremath{\mathrm{R}_{\odot}}}                  

\newcommand{\teff}{\ensuremath{\mathit{T}_{\rm eff}}}                

\usepackage{amsmath}	        
  
\newcommand{\fcbm}{\ensuremath{f_\mathrm{CBM}}}                             
\newcommand{\mco}{\ensuremath{M_\mathrm{CO}}}                             
\newcommand{\aov}{\ensuremath{\alpha_\mathrm{ov}}}                             

\definecolor{amethyst}{rgb}{0.6, 0.4, 0.8}

\definecolor{orange}{rgb}{0.726, 0.015, 0.015}

\title[GW190521 from first stellar generations]{Is GW190521 the merger of black holes from the first stellar generations?}

\author[E. Farrell et al.]{Eoin Farrell$^{1}$\thanks{E-mail: efarrel4@tcd.ie},
Jose H. Groh$^{1}$,
Raphael Hirschi$^{2,4}$,
Laura Murphy$^{1}$,
\newauthor
Etienne Kaiser$^{2}$,
Sylvia Ekstr\"om$^{3}$,
Cyril Georgy$^{3}$,
Georges Meynet$^{3}$
\\
$^{1}$School of Physics, Trinity College Dublin, The University of Dublin, Dublin, Ireland\\
$^{2}$Astrophysics Group, Keele University, Keele, Staffordshire ST5 5BG, UK\\
$^{3}$Geneva Observatory, University of Geneva, Chemin des Maillettes 51, 1290 Sauverny, Switzerland\\
$^{4}$Institute for the Physics and Mathematice of the Universe (WPI), University of Tokyo, 5-1-5 Kashiwanoha, Kashiwa 277-8583, Japan
}

\date{Accepted XXX. Received YYY; in original form ZZZ}

\pubyear{2020}

\begin{document}
\label{firstpage}
\pagerange{\pageref{firstpage}--\pageref{lastpage}}
\maketitle

\begin{abstract}
GW190521 challenges our understanding of the late-stage evolution of massive stars and the effects of the pair-instability in particular. We discuss the possibility that stars at low or zero metallicity could retain most of their 
hydrogen envelope until the pre-supernova stage, avoid the pulsational pair-instability regime and produce a black hole with a mass in the mass gap by fallback. We present a series of new stellar evolution models at zero and low metallicity computed with the Geneva and MESA stellar evolution codes and compare to existing grids of models. Models with a metallicity in the range 0 -- 0.0004 have three properties which favour higher BH masses. These are (i) lower mass-loss rates during the post-MS phase, (ii) a more compact star disfavouring binary interaction and (iii) possible H-He shell interactions which lower the CO core mass. We conclude that it is possible that GW190521 may be the merger of black holes produced directly by massive stars from the first stellar generations. Our models indicate BH masses up to 70-75~\msun. Uncertainties related to convective mixing, mass loss, H-He shell interactions and pair-instability pulsations may increase this limit to $\sim 85 \msun$.
\end{abstract}


\section{Introduction} \label{intro}


\begin{figure*}\centering
\includegraphics[width=0.82\linewidth]{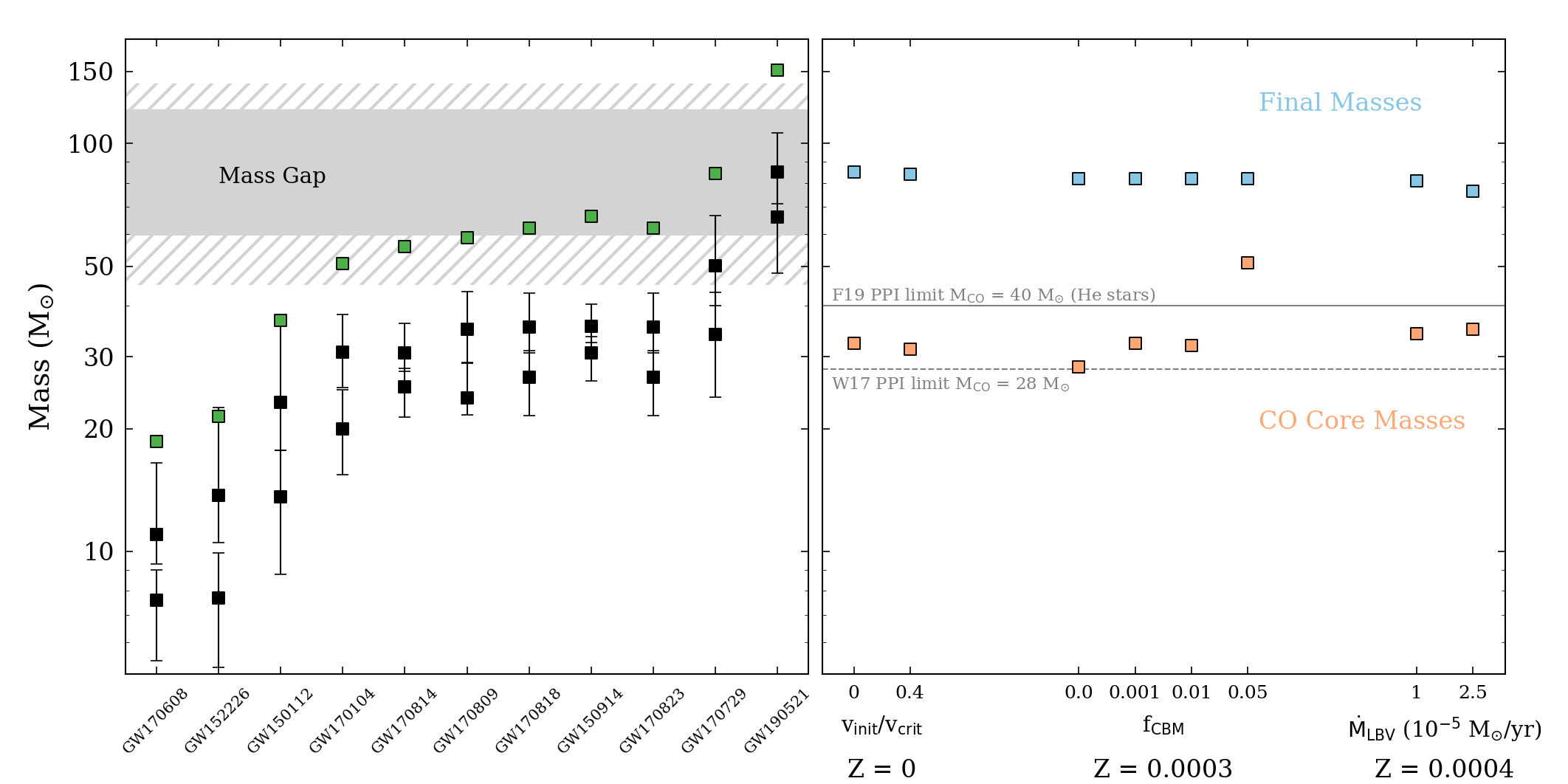}
\caption{\textit{Left panel:} Pre-merger and final BH masses from LIGO/Virgo observations in O1/O2 with GW190521 and the predicted region of the mass gap due to pair-instability. \textit{Right panel:} Final masses (blue) and CO core masses (red) of selected $85 \msun$ models listed in Table \ref{table1}. We also include the maximum CO core mass found by \citet{woosley17} that avoids any pulsations due to pair-instability.}
\label{ligo_figure}
\end{figure*}

The binary black hole merger GW190521 reported by the LIGO VIRGO Collaboration \citep{ligo20report, ligo20implications} contains unusually high component masses of $85^{+21}_{-14}$ and $66^{+17}_{-18} \msun$. These black hole masses lie within the mass gap predicted by standard (pulsational) pair-instability supernova theory. In this Letter we investigate the possibility that  stars at low or zero metallicity could retain most of their hydrogen envelope until the pre-supernova stage, avoid the pulsational pair-instability regime and produce a black hole with a mass in the pair-instability mass gap. 

In stars with CO core masses, $M_{\rm CO} \gtrsim 30 \msun$, the late nuclear burning phases are expected to be interrupted by the production of electron-positron pairs in the core \citep{fowler64, rakavy67}. For stars with CO core masses of $30 \msun \lesssim M_{\rm CO} \lesssim 60 \msun$, this can result in a series of energetic pulses followed by a collapse to a BH called a pulsational pair instability supernovae (PPISN) \citep{chatzopoulos12, chen14, woosley17, marchant19, leung19}. For $60 \msun \lesssim M_{\rm CO} \lesssim 120 \msun$, pair creation can result in a complete disruption of the star in a pair-instability supernova (PISN), leaving behind no remnant \citep{glatzel85, fryer01, umeda02, kasen11}. For even higher \mco, energy losses due to photo-disintegration are expected to result in a direct collapse to a BH \citep{fowler64, ober83, heger03, woosley07}. The combined effect of PPI and PI is predicted to produce a gap in the BH birth mass distribution between $\sim 55 - 130 \msun$ \citep{heger03, belczynski16, woosley19, giacobbo18}.
The exact boundaries of the mass gap are uncertain due to uncertainties in stellar evolution, core-collapse supernovae, PPISNe and PISNe \citep{woosley17, mapelli20, marchant19, farmer19, stevenson19, renzo20b}. \citet{farmer19} found that the lower boundary of the mass gap is quite robust against uncertainties in the metallicity ($\sim 3 \msun$), internal mixing ($\sim 1 \msun$) and stellar wind mass loss ($\sim 4 \msun$). However, they found that varying the $^{12}C (\alpha, \gamma) ^{16}O$ reaction rate within $1 \sigma$ uncertainties shifts the location of the lower-boundary of the mass gap between 40 and 56 \msun. \citet{vanSon20} investigated the possibility of super-Eddington accretion forming BHs in the mass gap, however they found no BBH with a combined mass $> 100 \msun$. Additionally, \citet{marchant20} investigated the impact of stellar rotation on the location of the mass gap and found that the lower boundary may be shifted upwards by 4 - 15\% depending on the efficiency of angular momentum transport. The boundaries of the pair-instability mass gap have also been proposed as a mechanism to place constraints on nuclear reaction rates \citep{farmer20}, particle physics \citep{croon20} and in cosmological studies \citep{farr19}.

Based on the BH mass function predicted by PPISNe and PISNe, the observation of a pre-merger $\sim 85 \msun$ BH as in GW190521 is unexpected. Several possibilities to create black holes with the reported mass are presented in previous works. The BH could form as a result of hierarchical mergers in dense stellar clusters, i.e. it is the result of the prior merger of two or more other BHs \citep[e.g.][]{miller02, gerosa17, fishbach17, rodriguez19, RomeroShaw20, Gayathri20, fragione20}. Other possible explanations include a stellar merger between a post-main sequence star and a main sequence binary companion \citep{spera19, diCarlo19}, a primordial origin \citet{DeLuca20}, different assumptions for stellar wind mass loss \citet{Belczynski20_highz}, Population III stars in binary systems \citep{Kinugawa20, Kinugawa20b, Tanikawa20}, an alternative prior in the gravitational wave analysis \citep{Fishbach20} and modifications to the Standard Model of particle physics \citep{sakstein20}. \citet{ligo20implications} found alternative explanations for the source of GW190521 to be highly unlikely, including a strongly gravitationally lensed merger or a highly eccentric merger. Given the widely predicted existence of the mass gap and the apparent robustness of the boundary of the gap with respect to uncertainties in stellar evolution models, can a single star produce a BH remnant with a mass around $85 \msun$?


\section{Stellar Evolution Models}
We present a series of new stellar evolution models computed with the Geneva Stellar Evolution code, \textsc{genec} (\citealt{ekstrom12}; Murphy et al. in prep) and with \textsc{mesa} \citep[r10398,][]{paxton11, paxton13, paxton15}. We also discuss the results from existing GENEC model grids (\citealt{ekstrom12,georgy13,groh19}).
Except where otherwise stated, the input physics for the \textsc{genec} and \textsc{mesa} models are similar to those described in \citet{ekstrom12} and \citet{choi16}, respectively. In our \textsc{mesa} models, we use the Ledoux criterion for convection with an exponential overshooting parameterised by \fcbm, while in our \textsc{genec} models, we use the Schwarzschild criterion with step-overshooting parameterised by \aov. In most models, we compute the evolution until at least the end of central C burning. For some \textsc{genec} rotating models, the computation is stopped at the end of He burning due to convergence difficulties. We define the CO core mass as the region where the helium abundance $Y < 0.01$ at the end of the evolution. The outputs from our models are summarised in Table \ref{table1}.

Figure \ref{ligo_figure} compares the LIGO binary black hole (BBH) masses \citep{ligo19o1o2} with the final masses and CO core masses of our models. The 85~\msun\ models with $Z$ in the range 0 to 0.0004 have final masses ranging from 76 to 85~\msun\ and CO core masses ranging from 28 to 51~\msun. In this metallicity range, the final mass depends on assumptions about convective boundary mixing and post-MS mass loss. Not surprisingly, the model with the lowest amount of convective boundary mixing (\fcbm = 0 and with the Ledoux criterion) produces the lowest CO core mass of 28~\msun. Increasing convective boundary mixing tends to produce higher CO core masses, however this depends on whether H-He shell interactions modify the convective core mass during Helium burning. For instance, H-He shell interactions impact the model with $\fcbm = 0.01$ at $Z = 0.0003$ so that despite the larger overshooting, its final CO core mass is lower than the model with $\fcbm = 0.001$.

\begin{figure}\centering
\includegraphics[width=0.93\linewidth]{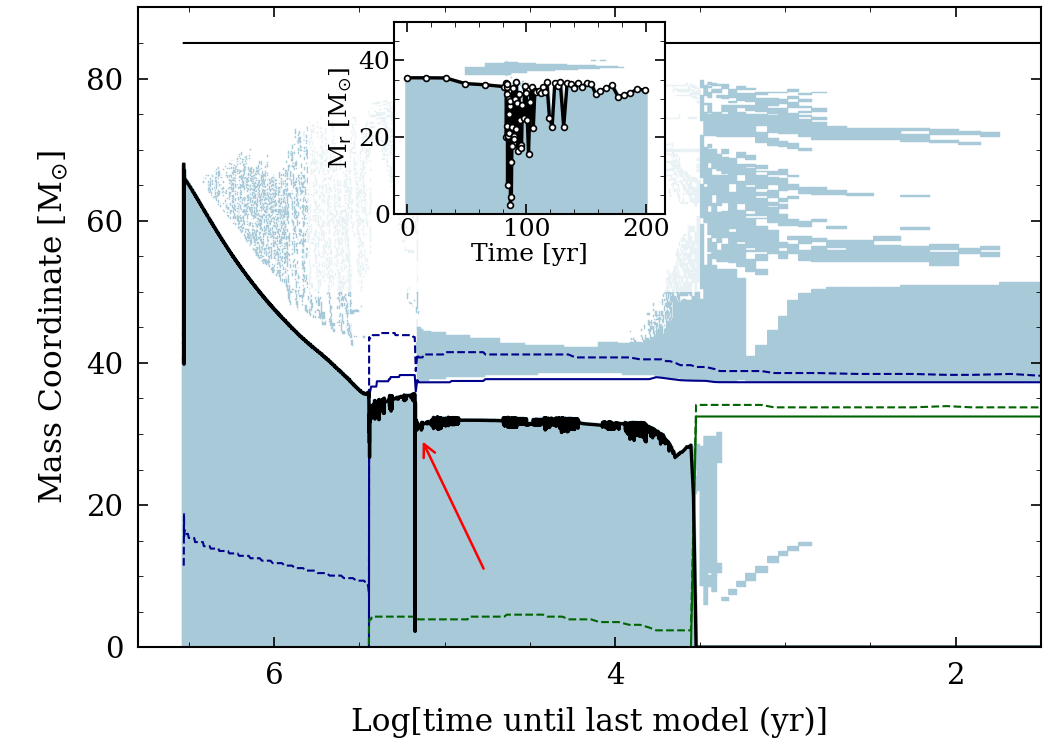}
\caption{Kippenhahn diagram of a GENEC non-rotating 85 \msun\ model at $Z=0$. Solid (dashed) lines correspond to the peak (100 erg/g/s) of the energy generation rate for H burning (blue) and He burning (green). The red arrow indicates the H-He shell interaction. An inset is included at the top of the figure to show that the interaction is resolved, where white circles indicate each timestep.}
\label{kipp_figure}
\end{figure}

H-He shell interactions are an interesting possibility to reduce the final CO core masses of massive stars at low and zero $Z$ \citep{ekstrom08,clarkson20}. This is relevant as it may allow a star to avoid the pulsational-pair instability regime, depending on initial mass and metallicity. To demonstrate this, we plot the Kippenhahn diagram of the evolution of our non-rotating $85 \msun$  $Z = 0$ stellar model (Fig. \ref{kipp_figure}). As expected, the convective core mass decreases during the MS evolution and increases following the onset of He-burning. However, shortly after the beginning of He-burning, the H-shell burning region becomes convective. This causes the convective core mass to decrease by $\sim 5 \msun$ (inset plot in Fig.~\ref{kipp_figure}) and prevents any subsequent increase as the star evolves to the end of He-burning.

\begin{figure}\centering
\includegraphics[width=0.85\linewidth]{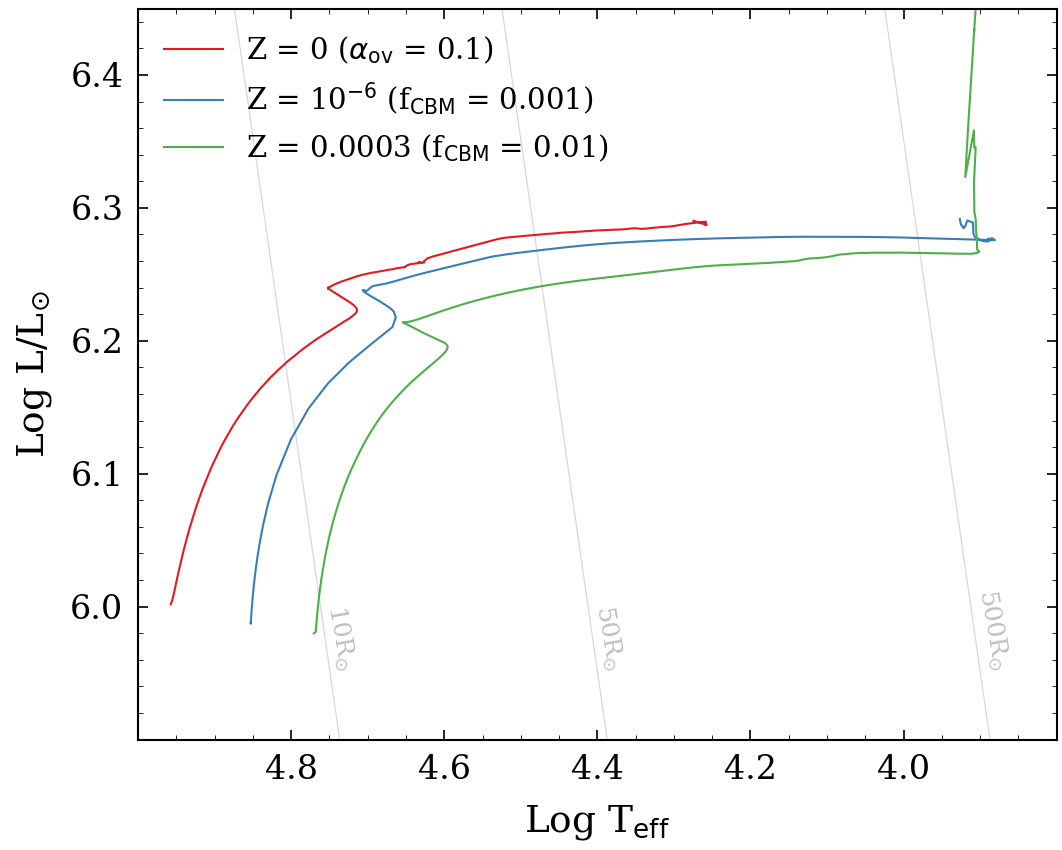}
\caption{Evolutionary tracks of selected $85 ~\msun$ models in the Hertzsprung-Russell diagram with $Z = 0$, $Z = 10^{-6}$ and $Z = 0.0003$.}
\label{hrd_figure}
\end{figure}

Figure \ref{hrd_figure} shows the evolutionary tracks in the Hertzsprung-Russell diagram of three $85 \msun$ models with metallicities of $Z = 0$, 10$^{-6}$ and 0.0003. The qualitative evolution during the MS is similar for all models. The location of the zero-age main sequence moves to higher \teff\ and luminosity with decreasing metallicity due the lower CNO abundances in the core. The post-MS evolution is affected in a similar way by the metallicity. At lower metallicites, a lower CNO abundance in the hydrogen-burning shell favours a more compact envelope and a higher \teff. This trend continues until the pre-supernova stage, so that the maximum radii that the models reach are 142, 672 and 794 \rsun\ for $Z = 0$, 10$^{-6}$ and 0.0003 respectively.

Previous works have focused on the context of producing BHs in close binary systems that could easily merge in the Hubble time and as a result assume that the entire H envelope will be lost to some combination of stellar winds, LBV eruptions or binary interaction \citep[e.g.][]{farmer19}. As a result, they focus on the evolution and deaths of helium stars \citep[e.g.][]{woosley19}. For single stars with hydrogen envelopes, a maximum BH mass of 60-65~\msun\ has been suggested for non-rotating models \citep{woosley17, mapelli20, spera17}. In their models, strong mass loss of the higher mass models coupled with higher core masses prevented the formation of higher mass BHs. Rotating models were found to have lower maximum BH masses. The models presented in this paper indicate black hole masses of up to 70 - 75 \msun, and possibly up to 85 \msun depending on uncertainties related to convective mixing, mass loss, H-He shell interactions and pair-instability pulsations. Our models leave open the possibility of a mass gap above $85 \msun$. To properly infer the actual limits of the pair instability mass gap based on these models, we would need to compute a large grid of models with different initial masses, rotation rates and metallicities. We defer this to future work.

\section{Implications for black hole masses from the first stellar generations}
Our models with $Z=0$ to 0.0004 have three properties which favour higher BH masses as compared to higher metallicity models. These are (i) lower mass-loss rates, in particular during the post-MS phase, (ii) possible H-He shell interactions which lower the CO core mass and (iii) a more compact star disfavouring binary interaction.

\begin{table}
\centering
\caption{Summary of our stellar evolution models. CBM refers to the free parameter regulating convective boundary mixing.}
\label{table1}
\begin{tabular}{rrrrrrr}
\hline
$Z$        & $M_{\rm zams}$  & CBM & Mass lost & $M^{\rm tot}_{\rm final}$ & $M^{\rm CO}_{\rm final}$ & $R_{\rm max}$\\
         & \msun\          & \aov /\fcbm     &      \msun       & \msun\          & \msun\              & \rsun\     \\
\hline
\multicolumn{7}{c}{Standard GENEC non-rotating models (\aov\ value given for CBM)}\\
\hline
0 & 60  & $0.1$    & 0.0 & 60.0  & 24.0 & 35   \\
0 & 85  & $0.1$    & 0.0 & 85.0  & 32.4 & 142  \\
0 & 120 & $0.1$    & 0.0 & 120.0 & 54.4 & 219  \\
\hline
\multicolumn{7}{c}{Standard GENEC rotating models (v = 0.4 $v_{\rm crit}$)}\\
\hline
0 & 60  & $0.1$    & 0.3 & 59.7  & 20.9 & 56   \\
0 & 85  & $0.1$    & 1.0 & 84.0  & 31.3 & 90   \\
0 & 120 & $0.1$    & 3.5 & 116.5 & 56.4 & 107  \\
\hline
\multicolumn{7}{c}{MESA models (\fcbm\ value given for CBM)}\\
\hline
$10^{-6}$   & 85   & 0.001   & 0.30  & 84.7  & 34.4  & 794         \\
0.0003 & 85 & 0.0   & 3.0 & 82   & 28.3 & 766       \\
0.0003 & 85 & 0.001 & 3.2 & 81.7 & 32.3 & 1169      \\
0.0003 & 85 & 0.01  & 3.0 & 82   & 32.0 & 672       \\
0.0003 & 85 & 0.05  & 7.0 & 78   & 51.0 & 984       \\
\hline
\end{tabular}

\end{table}

\subsection{Lower Mass Loss During the Evolution}
The amount of mass that a star retains until the pre-supernova stage depends strongly on its metallicity \citep[e.g.,][]{groh19}. This is a result of the strong dependence of mass loss from radiative-driven winds on metallicity \citep{vink01}. For solar metallicity stars, the time-averaged mass-loss rate during the LBV phase and the presence of surface magnetic fields are important factors that determine the final BH mass of massive stars, which can range from 35 to 71~\msun\ for an 85~\msun\ star \citep{groh19lbv}. At low metallicity, mass loss by stellar winds during the main-sequence phase becomes very low. Our 85~\msun\ models at $Z=0.0003$ lose only 1.5 \msun\ during the MS assuming the \citet{vink01} prescription. Further mass loss occurs during the post-MS and is strongly dependent on how cool the surface becomes. Our $Z=0.0003$ MESA models stay hot and lose 1.5~\msun\ during the post-MS, while our GENEC models can become spectroscopically similar to LBVs \citep{Groh_60msun_2014}. As a result, they may lose significantly more mass at that stage (7.5~\msun\ for $\dot{M}_{\rm LBV, max}$ = 2.5x10$^{-5} \msun$/yr),  even at low metallicity \citep{SmithOwocki06, allan20}.

At zero metallicity, radiatively driven mass loss becomes negligible throughout the evolution \citep{Krticka2006}, although for fast rotating stars there can be some small mass loss if the critical rotation limit is reached. Zero or negligible mass loss has been customarily used in stellar evolution grids at zero metallicity such as \citet{Marigo2001,ekstrom08,Yoon2012z=0,windhorst19} and Murphy et al. 2020, in prep. As such, our zero-metallicity models retain most of their mass until core collapse. There is little observational constraints for mass-loss rates at these extremely low-$Z$ values, in particular for the post-MS stages, and we should regard our assumptions about mass-loss rates as highly uncertain. Uncertainties related to mass-loss rates may affect both the final mass, the CO core mass and the maximum radius.

\subsection{Possibility of H-He Shell Interactions}
Some of our models at low/zero metallicity experience strong H-He shell interactions (Fig. \ref{kipp_figure}). This behaviour has been seen in previous low metallicity stellar evolution models \citep[e.g.][]{chieffi04, ekstrom08, ritter18, clarkson20}. During He-burning, a low or zero abundance of CNO elements in the H-burning shell favours a bluer star which increases the likelihood of the H-burning region becoming convective and subsequently reducing the convective core mass. In models with $Z = 0$, diffusion of C from the He-burning core to the H-burning shell can trigger a strong CNO cycle boost, make the shell convective and lead to H-He shell interactions. By comparing the \textsc{genec} models for metallicites of $Z$ = 0.0004, 0.002 and 0.014, \citet{groh19} discuss that the occurrence of H-He shell interactions may be favoured at lower metallicities. \citet{clarkson20} find different types of H-He shell interactions that occur at different times during the evolution. Some of these interactions, particularly during the late stages, may dramatically reduce the CO core mass and allow the star to avoid the pulsational pair instability regime. We encourage further work on the effects of convective boundary mixing and rotation on H-He shell interactions as this is crucial for understanding the fate of massive stars at low and zero metallicity.

Some of our models assume a relatively low amount of convective overshooting. The extent and implementation of convective overshooting in stellar models has a large impact on the mass of the He and CO cores \citep[e.g.][]{kaiser20}. Three-dimensional models of lower mass stars favour the existence of such mixing at convective boundaries \citep[e.g.][]{cristini17}, although it is still unclear how it is affected by other parameters such as mass and metallicity. In addition, for stars of initial mass  $7 < M_{\rm init} < 25 \msun$ a high value of \fcbm\ is favoured (Martinet et al. 2020, in prep) as well as for masses of $\sim 35 \msun$ \citep{higgins19}. However, these constraints are for core-H burning stars. The value of \fcbm\ is not as well constrained for other burning phases or for stars of $\sim 85 \msun$ which have different internal structures to $\sim 15 \msun$ stars and larger core mass ratios.

\subsection{Smaller Radius disfavours Binary Interaction}
Zero-metallicity models favour the retention of the H-envelope in binary systems because they are more compact than higher metallicity stars. For example, the maximum radius of our $85 \msun$ rotating model at Z = 0 is $R_{\rm max} = 142 \rsun$, as compared to $952 \rsun$ at Z = 0.0004 and $815 \rsun$ at Z = 0.014. The radius of stellar models at these masses depends greatly on the assumptions for convection in the envelope \citep[e.g.][]{grafener12a,jiang18}. Additionally, the radius is strongly impacted by uncertainties related to the chemical abundance profile in the envelope \citep{Farrell20}, which is impacted by the properties of mixing \citep[e.g.][]{schootemeijer19}. The size and interaction of convective shells above the core during the MS and between the MS and He-burning greatly affect the radius of the star during He-burning. If these processes result in hydrogen being mixed into the H-shell burning region, the star will remain more compact for longer during He-burning.

Binary interactions may also provide a mechanism to produce a pre-supernova structure with a high hydrogen envelope mass \citep[e.g.][]{Justham14}. Mass gainers or products of mergers during the post-MS that do not fully rejuvenate could have low core masses and large envelope masses, potentially avoiding the PPI regime and collapsing to a black hole with the H envelope falling back onto the BH \citep{spera19, diCarlo19}.

\subsection{Pulsational Pair-Instability}
Models suggest that stars with a CO core mass of $\gtrsim 28 \msun$ will undergo pair-instability driven pulsation during their final stages \citep{woosley17}. For example, \citet{woosley17} present a model (T80D) with a final mass of 80 \msun\ and a CO core mass of 32.6 \msun\ that, due to pulsations, will produce a final BH mass of 34.9\,\msun . The exact value of the maximum CO core mass of this boundary that will avoid the pair-instability is uncertain \citep[e.g.][]{woosley17, farmer19, marchant19} and effects related to convective boundary mixing, stellar winds and the $^{12}C (\alpha, \gamma) ^{16}O$ reaction rate may increase this value. Our 60 \msun\ models with Z = 0 have CO core masses between 21 and 24 \msun. Most of our 85\msun\ models are just above this strict limit with CO core masses of 31 -- 35 \msun.  We computed a test model with no convective boundary mixing that finishes with a CO core mass of 28 \msun. By interpolating between our 60 and 85 \msun\ models, we compute that a 72 \msun\ model will have a final CO core mass of 28 \msun\ under the standard assumptions for convection in the GENEC models.

For a pulse of a given energy, the amount of mass that a star loses depends on the binding energy of the envelope. More compact, hotter stars are less likely to lose their entire H envelope compared to extended envelopes, such as in RSGs. For this reason, $Z$ = 0 models are favoured to retain large masses as they remain compact until the end of their evolution. \citet{farmer19} find a CO core mass limit for the onset of PPI of $\sim 40 \msun$ for highly compact helium stars. Since our models are hydrogen rich, with a lower binding energy than helium stars, it is unclear if this limit would apply to our 85~\msun\ models. Further studies could investigate the impact of the uncertainties discussed by \citet{farmer19}, such as the $^{12}C (\alpha, \gamma) ^{16}O$ reaction rate, in hydrogen-rich models that are blue and relatively compact, such as our $Z=0$ models. If the pulses are not present and/or do not remove the H envelope, this may allow the formation of $85~\msun$ BHs.

\section{Impacts for Binary Black Hole Mergers} 
Due to their lower mass-loss rates, smaller 
radii and the possibility of H-He shell interactions that reduce the CO core mass, stars in the first stellar generations are 
ideal candidates to produce BHs in the mass gap such as GW190521, with masses of 70 -- 75\, \msun\ . In order to produce a BBH merger observable by LIGO/Virgo, such a BH would need to be in a close binary system. Due to uncertainties in the evolution of massive stars and in how these stars behave in binary systems, it is difficult to perfectly constrain the possible evolutionary pathways that would lead to a system. 
Despite their large H-envelope mass, our models at $Z=0$ expand only to radii $\sim 100 \rsun$.
If the star has a binary companion and avoids Roche-Lobe overflow, the merging timescale would likely exceed the Hubble time. However, if the orbital separation were to reduce after the more massive star dies (e.g. due to a common envelope phase) this may reduce the merging timescale. Alternatively, if the BH is in a dense stellar cluster, it could dynamically capture a companion and form a close binary system \citep[e.g.][]{Sigurdsson93, Portegies00, Downing10, Rodriguez16}. We leave the details of the binary evolution scenario or dynamical capture to future work \citep[e.g.][]{Belczynski20_binary}.

\medskip
\noindent \textbf{Acknowledgements}: EF, JHG and LM thank the Irish Research Council for funding. JHG thanks Isidoros and Giorgos for providing useful resources. This article is based upon work from the ``ChETEC'' COST Action (CA16117). RH acknowledges support from the IReNA AccelNet Network of Networks (NSF Grant No. OISE-1927130) and from the World Premier International Research Centre Initiative. CG has received funding from the ERC (grant No 833925).

\medskip
\noindent \textbf{Data availability}: The derived data generated in this research will be shared on reasonable request to the corresponding author.

\bibliographystyle{mnras}
\bibliography{refs}

\label{lastpage}
\end{document}